\documentstyle[editedvolume,psfig]{crckapb} 

\begin{opening}
\title{Extended gas in interacting systems}
\subtitle{}

\author{F. Combes}
\institute{Observatoire de Paris, DEMIRM\\
           61 Av. de l'Observatoire, F-75 014 Paris, France}

\end{opening}

\runningtitle{Extended gas in interacting systems}

\begin{document}

\begin{abstract}
HI observations have revealed large gaseous extensions in interacting
and merging systems. The interstellar gas is obviously dragged out
in tidal tails during an encounter, and the percentage of HI in the tails
increases with the merging stage. However, the opposite is true for
the molecular gas, which is observed highly concentrated towards 
the nuclei of interacting galaxies, amounting to a significant fraction
of the dynamical mass. Statistically, there appears to be more gas
{\it observed} in interacting galaxies than in normal, isolated ones.
As N-body simulations show, the gas 
is driven inwards in the interaction process by the strong
gravity torques, before being consumed through star formation
in the triggered starbursts. We review here all observations that
could bring more knowledge about the state of the gas in the outer
parts of galaxies, and about accretion processes. The link 
with the observations of the Ly$\alpha$ absorbers at low and high
redshifts is discussed.
\end{abstract}

\section{Fate of the gas in interacting systems}

One of the main striking features about interacting and merging
galaxies, is the presence of large tidal tails of matter dragged out of
the galaxies; recent VLA maps have revealed huge HI extensions
with respect to the optical systems, as though most of the neutral
gas was splashed all around. Yun et al. (1993) have found large quantities
of HI all around the M81/M82/NGC 3077 system, and Hibbard (1995) shows
in his thesis an evolving sequence of interacting/merging galaxies,
where the HI extensions are conspicuous. More precisely, the percentage 
of the total HI found in the tails/extensions is increasing with
the merging stage, from 20\% in the M81 system, to 80\% in the
merger remnant NGC 7252. Of course, this can be explained by the tidal 
potential in the frame of the target growing as r$^2$ as a function of 
the target radius, and because matter in the outer parts is less bound 
to the galaxies. But this gives a wrong idea of the fate of the gas 
in interacting systems. In fact, with all probability, the gas 
dragged out remains bound to the system, and will rain back onto the merger
remnant, after some billion years. Already, Hibbard (1995) shows that the
gas at the bottom of the tails in NGC 7252 is infalling. This progressive
infall will take place through phase wrapping, and shells and loops
will form. 

What is seen in the molecular phase is just the contrary: apparently 
large H$_2$ concentrations pile up at the galaxy nuclei in interacting
systems. Up to 50\% of the dynamical mass could be under
the form of molecular hydrogen in merging systems (Scoville et al. 1991).
Some ultra-luminous infrared galaxies, which are also mergers and
starbursts possess 10$^{10}$ M$_\odot$ of H$_2$ gas, about 10 times more
than in their spiral precursors. There exist some hints that there is
more gas in interacting galaxies (Braine \& Combes 1993). Although
the H$_2$/CO conversion ratio is not well known for this peculiar
objects, there is evidence for denser gas in these systems,
through the HCN/CO ratio (Solomon et al. 1997).
In summary, the observations suggest that the HI gas is dragged outwards,
while the H$_2$ gas is driven inwards, to be consumed in star formation.
In fact these two tracers (HI and CO) shed light on two aspects of
the same gas component. 

The global result is  that most of the dissipative component loses
angular momentum, and falls inwards towards the nucleus of the
merger, although part of it is heated
in shocks and pass in the coronal phase (seen in X-rays).

\section{Extension of gas in normal galaxies}

\subsection{ Spiral galaxies}
 It is well known that the HI gas is extending sometimes 
much farther than the stars in spiral galactic disks, and
they are precious tools to probe the rotation curve and the
presence of dark matter. However, large extensions such as in NGC
628, where R$_{HI}$ = 5 R$_{25}$ for instance, are quite exceptional
(Kamphuis \& Briggs 1992); only about 10\% have  R$_{HI}$ larger
than 2.5  R$_{25}$ (Huchtmeier \& Seiradakis 1985).
In a sample of about 100 galaxies, chosen to search precisely 
for extended HI disks, Broeils (1992) did not find many extended
gaseous disks. In fact, only regular galaxies were selected, to 
be able to exploit the rotation curves, and therefore no strongly
interacting galaxies are included. The HI diameter, defined at a surface
density of 1 M$_\odot$ pc$^{-2}$, is about twice the optical diameter
D$_{25}$ (Broeils \& van Woerden 1994). The HI-to-optical-diameter
ratio does not depend on morphological type or luminosity (fig. 1), but
there is a strong correlation indicating that M$_{HI} \propto$ D$_{HI}^2$,
which means that HI surface density, averaged over the whole HI
disk, is constant from galaxy to galaxy, and all over the Hubble sequence
(Broeils \& Rhee 1997).
 Besides, there is a strong dependence of the ratio  M$_{HI}$/M$_{tot}$ with
type, which confirms that the percentage of gas is larger in 
small late-type galaxies, and decreases all over the Hubble sequence
(which is also a mass sequence).

\begin{figure}
\psfig{figure=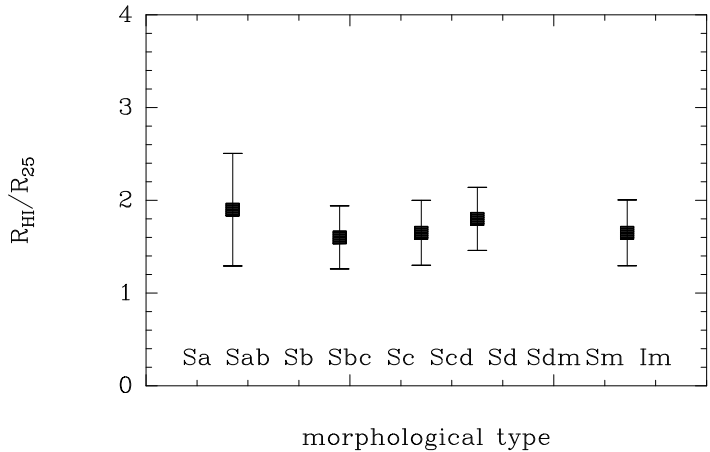,bbllx=50mm,bblly=4mm,bburx=180mm,bbury=240mm,width=12.5cm,angle=-90}
\caption{ Ratio of HI radius (defined by a surface density of
1 M$_\odot$ pc$_{-2}$) to optical radius R$_{25}$ as a function of morphological type,
from Broeils \& Rhee (1997). The error bars are 1 $\sigma$ dispersion.
}
\end{figure}

\subsection{ Dwarf irregular galaxies}

Since dwarf irregular galaxies are particularly rich in HI gas, it was first
thought that there might exist big HI envelopes around these objects, or
even a large number of isolated gas clouds still waiting to form their first
stars. It turned out however that big gas extensions, such as in DDO154
are very uncommon (e.g. Hoffman et al. 1996), and that the HI-to-optical-radius
ratio is very similar in dwarfs and in spirals (cf. fig. 2).

\begin{figure}
\psfig{figure=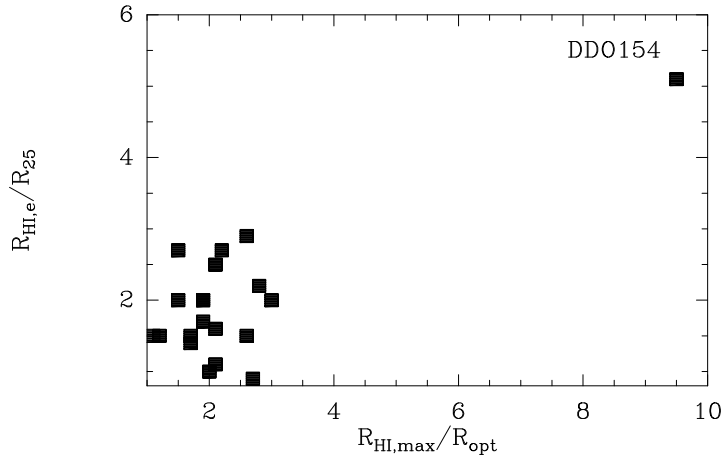,bbllx=50mm,bblly=4mm,bburx=180mm,bbury=240mm,width=12.5cm,angle=-90}
\caption{ The ratio of maximum HI radius to optical radius, and
the ratio of HI radius at 1/e of the peak flux to the optical
isophotal radius, for all resolved dwarfs, 
studied by Hoffman et al. (1996). Only DDO 154 is outstanding in its
HI size. }
\end{figure}

\subsection{ The ionization edge of the HI disk }

HI observations have shown that the neutral gas radial distribution
is much smoother than that of stars (exponential decrease), or molecular gas
as traced by CO (following more or less the blue luminosity). The HI surface density
is statistically falling as the dark matter surface density, determined from
rotation curves (Bosma 1981, Freeman 1993). This implies an HI column 
density falling as $1/r$ at large radii (because the gas layer is flaring
linearly with radius, the volumic density in fact is going as  $1/r^2$).
It would be interesting to observe
the HI until the edges of the neutral disk, to trace its physical condition and
also the distribution of the dark matter at large radii, but this is a hard task,
due to sensitivity problems. It has been done in a few cases only, in NGC 3198
(van Gorkom 1991) and M33 \& NGC 3344 (Corbelli et al. 1989). The results show
a sudden decrease from a column density of N$_{HI} = 2 \cdot 10^{19}$ cm$^{-2}$ to
 2 $\cdot 10^{18}$ cm$^{-2}$ in a very short length-scale of 2-4 kpc (of course limited by
the spatial resolution of the observations). This sharp edge has been 
interpreted in terms of the ionization front of the gas disk
(Corbelli \& Salpeter 1993): the data can be reproduced if the extragalactic
background, essentially from the quasar UV light, provides an 
ionization rate of $\xi \sim 2 \cdot 10^{-14}$ s$^{-1}$, a value corresponding
to the study of low-redshift Lyman-$\alpha$ absorption lines (Madau 1992).
  The conclusion is that the gas surface density is likely to continue to
decrease as $1/r$ at large distances, but it is difficult to see it.
It is even possible that the gas density in the outer parts is
highly underestimated, if it is clumpy down to very small scales,
and under a cold molecular phase (Pfenniger et al. 1994, Pfenniger \& Combes 
1994).
Large HI extensions, that are only seen when galaxies are interacting,
and molecular gas concentrations, 
could then reveal the presence of the clumps when they are stirred up
by tidal interactions, driven inwards and concentrated.

\subsection{ How to trace gas at large radii? }

Gas at large radii is difficult to trace, because of its low average
column density, its low temperature, low metallicity Z and dust content 
(Z is decreasing exponentially with radius, e.g. Smartt \& Rolleston 1997). 
Even at any radii,
the usual tracers are far from perfect: the HI line can be optically
thick (e.g. Burton 1992), CO emission can be absent by lack of
excitation for example (Adler et al 1991). Nelson et al. (1996)
claim to have detected cold dust at large radii in nearby galaxies
through its 100$\mu$ emission; but the very cold dust could be more easily
traced at 1.3mm. Recent continuum maps at this wavelength (which is
in the Rayleigh-Jeans domain), are still limited by sensitivity, but they 
reveal interesting radial distributions. In galaxies where the 
interstellar medium (ISM) is dominated by the molecular component, like NGC 891,
the 1.3mm flux radial distribution is superposable to the CO one 
(Gu\'elin et al. 1993), suggesting that the CO emission is directly proportional
to metallicity (like dust emission since S$_{1.3mm}$ is proportional to
the dust temperature, the column density of the gas, and the metallicity). 
On the contrary, in galaxies where the ISM is dominated by the 
atomic hydrogen, like NGC 4565, the 1.3mm continuum flux follows more
the HI emission, which is decreasing more slowly with radius than
the CO one (Neininger et al. 1996),
but still the dust emission is falling more rapidly than the HI, because
of its Z dependency. In spite of sensitivity and metallicity limitations,
dust emission could be one of the best way to trace extended cold
gas (e.g. Combes \& Pfenniger 1997).

\subsection{Stability of extended gaseous disks}
  If extended gaseous disks exist, their gravitational stability 
raises interesting problems. Since the HI gas at large radii is
usually observed lopsided (Richter \& Sancisi 1994), or with multiple spiral 
arms and large-scale instabilities, it must possess a minimum of self-gravity
(Pfenniger et al. 1994). As for small scale instabilities,
as well as the vertical equilibrium, constraints can be put on
the flattening of the dark matter haloes, which determines
its volumic density in the plane, as shown in fig. 3
(cf. Olling 1995, 1996, Becquaert \& Combes 1997).

\begin{figure}
\psfig{figure=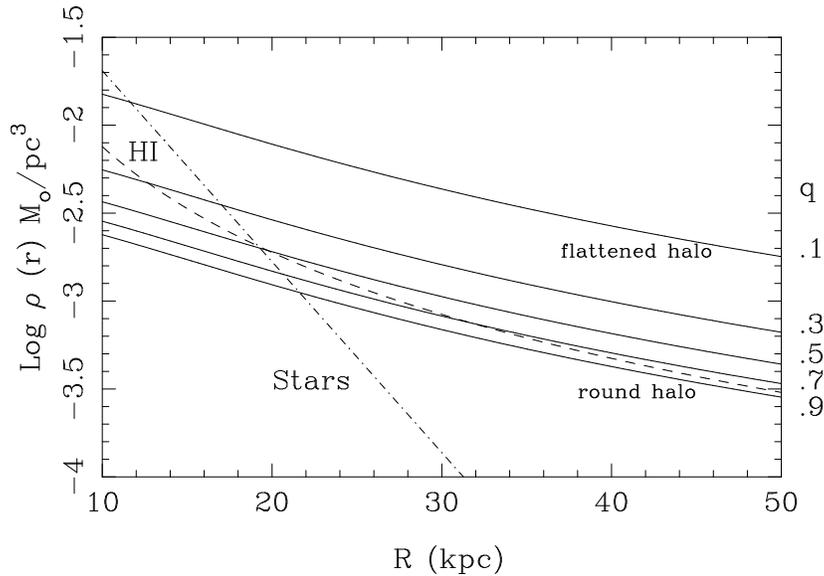,bbllx=20mm,bblly=4mm,bburx=210mm,bbury=280mm,width=12.5cm,angle=-90}
\caption{ Volumic density as a function of radius for the various mass components
in a typical galaxy like the Milky Way: exponential stellar disk (dot-dash),
flaring HI layer (dash), and isothermal spheroidal halo of flattening q (full lines).
}
\end{figure}

\section{Intergalactic medium}

\subsection{Galaxy clusters}

Through frequent tidal interactions or harassment (Moore et al. 1996),
and ram-pressure stripping (Cayatte et al. 1990),
the extended gaseous disks are truncated in rich clusters. Globally,
the HI surface density is conserved for spirals in clusters, only their
HI-to-optical-radius ratio is smaller (Cayatte et al. 1994), and since their
CO content is also normal (Casoli et al. 1996), their central gaseous disk appears
not too much perturbed, except for a small number of anaemics. The gas coming from
the outer parts of galaxies must be now under the form of X-ray emitting hot gas,
at the virial temperature of the cluster, and enriched in heavy elements
from the galaxies (Renzini 1997).

\subsection{Lyman-$\alpha$ absorbers}

The intergalactic medium is traced through Lyman-$\alpha$ absorption in front
of quasars, especially well sampled at $z=2$ from the ground. Space observations
have shown that the absorbers are still present until $z=0$, although with some
evolution in number density. The distribution of HI column densities derived
from these probes is a continuous power-law (slope $\sim$ -1.6) up to damped 
Ly-$\alpha$ values of 10$^{20}$-10$^{21}$ cm$^{-2}$, comparable to the outer parts of
galaxy disks. Between $z=2$ and $z=3$, most of the baryons are in the Lyman-$\alpha$ 
forest with N$_{HI}$ between  10$^{14}$ and 10$^{16}$ cm$^{-2}$, but the medium
is strongly photoionized, with the neutral fraction around 10$^{-6}$-10$^{-5}$
(the total column density is thus much higher). Typical sizes are 200kpc,
certainly in filamentary structures. Recent ionized helium detections
(Jacobsen et al. 1994, Davidsen et al. 1996) reveal that the most diffuse regions
of the intergalactic medium are filled with ionized gas, in other words
the Gunn-Peterson effect is detected in HeII.

What is less well known is the high end of the density spectrum, and
whether small clumps of dense molecular gas exist in the intergalactic medium.
 In the frame of the fractal model or hierarchy of gas structures proposed
for the ISM, physical mechanisms can explain the dynamical equilibrium
and the non star formation (Pfenniger \& Combes 1994). These structures,
down to the smallest H$_2$ clumps, could be formed at high redshift
(z $>$ 100), as soon as the cooling time is much smaller than the Hubble time.
The clumps will form before re-heating, and represent an alternative model
to solve the cooling catastrophy.

In any case, galaxies are not isolated objects, they are connected to
the gas reservoirs of the intergalactic medium, which contains most of the
baryons at $z=2$. This should be taken into account when considering
galaxy interactions. If the external gas has a flattened geometry, a big 
disk or ring can form in a merging sequence (like is observed in NGC520
for instance). The interaction produces compression, higher densities
and higher neutral fraction (which goes as the square of the density)
and the gas will become visible in the HI line.

\section{Conclusion}

There are various pieces of evidence for large gas reservoirs in the outer
parts of galaxies. This gas is driven inwards during galaxy interactions,
through the associated gravity torques. It can help to explain mergers and 
the triggered starbursts, permanent gas accretion in warps, polar rings, etc...
If the medium is clumpy and fractal, the gas mass could be underestimated
by factors 2-10, and this can have important cosmological consequences.

\section{References}


\noindent Adler D.S., Allen R.J., Lo K.Y.: 1991, ApJ 382, 475

\noindent Becquaert J-F., Combes F.: 1997, A\&A 345, 41

\noindent Bosma A.: 1981, AJ 86, 1971

\noindent Braine J., Combes F.: 1993, A\&A 269, 7

\noindent Broeils A.: 1992, PhD Thesis, Groningen

\noindent Broeils A., van Woerden H.: 1994, A\&AS 107, 129

\noindent Broeils A., Rhee M.H.: 1997, A\&A 324, 877

\noindent Burton W.B.: 1992, in "The Galactic Interstellar Medium", Saas-Fee
 Advanced Course 21, ed. D. Pfenniger \& P. Bartholdi, p. 1

\noindent Casoli F., Dickey J., Kazes I. et al.: 1996, A\&A 309, 43

\noindent Cayatte V., Balkowski C., van Gorkom J.H., Kotanyi C.: 1990, 
AJ 100, 604  

\noindent Cayatte V., Kotanyi C., Balkowski C., van Gorkom J.H.: 1994, 
AJ 107, 1003

\noindent Combes F., Pfenniger D.: 1997, A\&A 327, 453

\noindent Corbelli E., Salpeter E.E.: 1993, ApJ 419, 94 \& 104

\noindent Corbelli E., Schneider S.E., Salpeter E.E.: 1989, AJ 97, 390

\noindent Davidsen A.F., Kriss G.A., Zheng W.: 1996, Nature 380, 47

\noindent Freeman K.C.: 1993, in "Physics of nearby galaxies, Nature or Nurture?
ed. T.X. Thuan, C. Balkowski, Van J.T.T., Editions Fronti\`eres, Gif-sur-Yvette, p. 201

\noindent Gu\'elin M., Zylka R., Mezger P.G. et al.: 1993, A\&A 279, L37

\noindent Hibbard J.E.: 1995, PhD thesis, Columbia University

\noindent Hoffman G.L., Salpeter E.E., Farhat B. et al.: 1996, ApJS 105, 269

\noindent Huchtmeier W.K., Seiradakis J.H.: 1985, A\&A 143, 216

\noindent Jacobsen P. et al.: 1994, Nature 370, 35

\noindent Kamphuis J., Briggs F.: 1992, A\&A 253, 335

\noindent Madau P. 1992, ApJ 389, L1

\noindent Moore B., Katz N., Lake G.: 1996, ApJ 457, 455

\noindent Nelson A.E., Zaritsky D., Cutri R.M.: 1996, AAS, 189, 6706

\noindent Neininger N., Gu\'elin M., Garcia-Burillo S., Zylka R., 
Wielebinski R.: 1996, A\&A 310, 725

\noindent Olling R.P.: 1995, AJ 110, 591

\noindent Olling R.P.: 1996, AJ 112, 457

\noindent Pfenniger D., Combes F., Martinet L.: 1994, A\&A 285, 79

\noindent Pfenniger D., Combes F.: 1994, A\&A 285, 94

\noindent Renzini A.: 1997, ApJ 488, 35

\noindent Richter O-G., Sancisi R.: 1994, A\&A 290, L9

\noindent Smartt S.J., Rolleston W.R.: 1997, ApJ 481, L47

\noindent Scoville, N.Z., Sargent, A.I., Sanders, D.B., Soifer, B.T.: 1991 
       ApJ, 366, L5

\noindent Solomon P.M., Downes D., Radford S.J.E., Barrett J.W.: 1997, ApJ 478, 144

\noindent van Gorkom J.H.: 1991, in "Atoms, Ions and Molecules", ed. A.D.
Haschik, ASP Conf. Ser. 16, 1

\noindent  Yun M.S., Ho, P.T.P., Lo K.Y.: 1993, ApJ, 411, L17


\end{document}